\documentclass[pre,10pt,superscriptaddress,floatfix,showpacs,notitlepage]{revtex4-1}
\usepackage[utf8]{inputenc}  
\usepackage[T1]{fontenc}     
\usepackage[british]{babel}  
\usepackage[sc,osf]{mathpazo}
\usepackage[scaled=0.86]{berasans}  
\usepackage[colorlinks=true, citecolor=blue, urlcolor=blue]{hyperref}  
\usepackage{graphicx} 
\usepackage[babel]{microtype}  
\usepackage{amsmath,amssymb,amsthm,bm,amsfonts,mathrsfs,bbm} 

\begin{document}
\title{Implications of  Coupling in Quantum Thermodynamic Machines}

\author{George Thomas}
\email{georget@imsc.res.in}
\affiliation{Optics and Quantum Information Group, The Institute of Mathematical Sciences, HBNI,  C. I. T. Campus, Taramani, Chennai 600113, India.}

\author{Manik Banik}
\email{manik11ju@gmail.com}
\affiliation{Optics and Quantum Information Group, The Institute of Mathematical Sciences, HBNI,  C. I. T. Campus, Taramani, Chennai 600113, India.}

\author{Sibasish Ghosh}
\email{sibasish@imsc.res.in}
\affiliation{Optics and Quantum Information Group, The Institute of Mathematical Sciences, HBNI,  C. I. T. Campus, Taramani, Chennai 600113, India.}

\begin{abstract}
We study coupled quantum systems as the working media of thermodynamic machines. 
Under a suitable phase-space transformation, the coupled systems
can be expressed as a composition of independent subsystems. 
We find that for the coupled systems, the figures of merit, that is the efficiency for engine and the coefficient 
of performance for refrigerator, are bounded 
(both from above and from below) by the corresponding figures of merit of the independent subsystems.
We also show that the optimum work extractable from
a coupled system is upper bounded by the optimum work obtained from the uncoupled system, thereby 
showing that the quantum correlations do not help in optimal work extraction.
Further, we study two explicit examples; coupled spin-$1/2$  systems and coupled quantum oscillators with analogous 
interactions. Interestingly, for particular kind of interactions, the 
efficiency of the coupled oscillators outperforms that of the coupled spin-$1/2$ systems 
when they work as heat engines. However, for the same interaction, the coefficient of performance 
behaves in a reverse manner, while the systems work as the refrigerator.
Thus the same coupling can cause opposite effects in the figures of merit of heat engine and refrigerator. 
\end{abstract}
\maketitle

\section{Introduction}
Study of thermodynamics in quantum regime can reveal new aspects of fundamental interests. 
As an example, the statement of the second law of thermodynamics in the presence of an 
ancilla \cite{Horodecki'2013,Wehner'2014} or, when the system has coherence \cite{Lostaglio'2015(1),Lostaglio'2015(2)}, 
has been established in great details, from where the  classical version of the second 
law emerges under appropriate limits. The study of thermodynamics in quantum domain can be approached from different directions
such as information-theoretic point of view \cite{Landauer'1961,Bennett'1982,Lubkin'1987,Vedral'2009,Vedral'2011,Thomas'2012} 
or resource-theoretic aspect \cite{Spekkens'2013,Oppenheim'2013,Gour'2015}. Another 
important constituent, in this area of study, is the work extraction from quantum 
systems \cite{Aberg'2013,Popescu'2014,Aberg'2014,Acin'2014,Mukherjee'2016}. Besides these, analyzing 
different models of thermodynamic machines in the quantum domain can also provide new insight. 
Such a study helps us to understand the special behavior of thermodynamic quantities like work, heat, and 
efficiency in the quantum regime due to the presence non-classical features such as entanglement, 
quantum superposition, squeezing, etc.\cite{Scully'2003,Kosloff'2013,Lutz'2014}. 
The quantum heat devices can show interesting atypical behaviors such as  
exceeding Carnot limit \cite{Scully'2003,Lutz'2014} when they act as heat engines. But these apparent behaviors
 are found to be compatible with the second law of thermodynamics when  all the preparation costs are considered \cite{Zubairy2002}. Such machines 
also have practical importance in the realm of quantum computation and refrigeration of small 
systems \cite{Linden'2010}. 

The performance of coupled quantum systems as heat engines have 
been studied widely in recent past \cite{Kosloff'2002,Kosloff'2003,Zhang2007,Thomas'2011,Thomas'2014,Azimi'2014,Wang'2015}. 
Work and efficiency are  two important quantities to characterize the performance of a heat engine.
It has been shown that appropriate coupling can increase the efficiency of the system compared
 to the uncoupled one \cite{Thomas'2011}. In this work, we find an upper as well as a lower
 bound of the efficiency of the coupled system. We also show that
 the coupling and quantum correlations give
 no advantage to obtain optimum work when the 
 working medium consists of quantum systems with quadratic coupling (to be specified later).
 The generality of these results are shown by considering
 different heat cycles. 
 
 Further, we compare the performances of different 
coupled quantum systems when used as the working medium of a heat device. For this purpose, we 
consider two extreme cases: coupled spin-$1/2$ systems (finite dimension) and coupled quantum 
oscillators (infinite dimension) as the working media of  heat devices. 
In the case of Otto cycle, when there is no coupling, both the system has the same efficiency but 
the work output is higher for oscillator model. To compare the performance of the coupled 
systems, an analogous coupling for both the systems is taken. We consider that two spin-$1/2$ 
systems are coupled via Heisenberg XX or XY model of interaction \cite{Lieb1961,Takahashi'1999}. 
For harmonic oscillators, we take quadratic interaction in both positions and momenta, which is 
analogous to the Heisenberg exchange interaction in spin systems. For this interaction, the 
Hamiltonians for coupled spins and coupled oscillators are similar in terms of ladder operators.
The similar form of the free Hamiltonians for both of the systems is also ensured.  

Here it is important to note that very recently, the  performance of coupled harmonic oscillators 
as a heat engine is studied \cite{Wang'2015}. But, in contrast to the present work, there the
 interaction has been considered only between the position degrees of freedom of two oscillators. 
Furthermore, the authors in \cite{Wang'2015} have done the efficiency analysis for two different 
modes separately. But, the actual efficiency of the system has to be defined by the ratio of 
total work (done by both the modes) to the total heat which we analyze here. Therefore, our 
analysis provides a comprehensive picture of the efficiency of the coupled system. The main 
results of our work are as follows:
\begin{itemize}
\item[(i)] When the Hamiltonian of the coupled system (at all stages of the cycle) can be 
decoupled (as two independent modes) in some suitably chosen co-ordinate system, then the 
efficiency of the coupled system is bounded (both from above and below) by the efficiencies 
of the independent modes, provided both the modes work as engines (Section \ref{sec3A}).
\item[(ii)] The global efficiency (i.e. efficiency of coupled system) reaches the lower bound (mentioned in (i)) 
when the upper bound (mentioned in (i)) of the efficiency achieves Carnot efficiency. When one of the modes is not working as an engine, the global 
efficiency is upper bounded by the efficiency of the other mode (Section \ref{sec4A}).
\item[(iii)] For the case of the engine, we compare the efficiencies in two extreme cases 
(coupled oscillators and coupled spin-$1/2$ systems). Interestingly,  the efficiency of 
coupled oscillators outperforms the efficiency  obtained from coupled spins (Sections \ref{sec4A} and \ref{sec4B}).
 \item[(iv)] We have also shown that the optimal work extractable from a coupled system 
is upper bounded by the optimal work extractable from the uncoupled systems (Section \ref{sec3C}).
\end{itemize}
For meticulous comparison, we also consider coupled oscillators and coupled spins as 
the working medium of a refrigerator. The refrigeration cycle is similar to that of the 
heat engine. We find that:
\begin{itemize}
\item[(v)] Like the efficiency, the global coefficient of performance (COP) is bounded
 (both from above and below) by the COPs of the independent modes (Section \ref{sec5}).
\item[(vi)] Surprisingly, for similar interactions considered in the case of heat engine, 
the global COP of coupled spins is higher than that of the coupled oscillators, which is
 contrary to the behavior observed in the case of engines (Sections \ref{sec5A} and \ref{sec5B}). 
\end{itemize}

Organization of the paper goes as follows: In Section-\ref{sec2}, we introduce the Otto 
cycle and illustrate the performance of uncoupled spins and oscillators as the working
 substance. In  Section-\ref{sec3}, a general form of quadratic coupling in harmonic 
oscillators and their characteristics are discussed when they work as a heat engine.
 Further, we discuss similar forms of coupling found in spin systems, widely known as 
Heisenberg XY model. In Section-\ref{sec4}, we describe the performance of the engine 
for special cases. Performances of the systems as refrigerators is discussed in 
Section-\ref{sec5}. Section-\ref{sec6} is devoted to discussions and future possibilities.

\section{Quantum Otto cycle}\label{sec2}
Quantum Otto cycles are analogous to the classical Otto cycle, and the latter consists 
of two isochoric processes (work, $W=0$) and two adiabatic processes (heat, $Q=0$). 
When the working medium of  the Otto cycle is classical ideal gas, the efficiency of 
the system is written as $\eta=1-(V_1/V_2)^{\gamma-1}$, where $V_1$ and $V_2$ are initial 
and final volumes ($V_1< V_2$) of the adiabatic expansion process and $\gamma=C_p/C_v$, 
is the ratio of the specific heats \cite{Schroeder'2000}. Similar to the classical cycle,
 the quantum Otto cycle consists of two adiabatic processes and two thermalization 
processes \cite{Kieu'2004,Quan'2007}. The system exchanges heat with the bath during 
the thermalization processes and the work is  done when the system undergoes adiabatic 
processes. Work and heat are calculated from the change in mean energies, where mean 
energy, for a  system represented by the state $\rho$ and the Hamiltonian $H$,  is defined 
as ${\rm Tr} [\rho H]$.

Here, we consider a four-staged Otto cycle. As an example, harmonic oscillator as the working
medium of a  quantum Otto cycle is pictorially described in  Fig. \ref{cycle}. The four 
stages of the cycle are:
\begin{figure}[h!]
\begin{center}
\vspace{0.2cm}
\includegraphics[width=8cm]{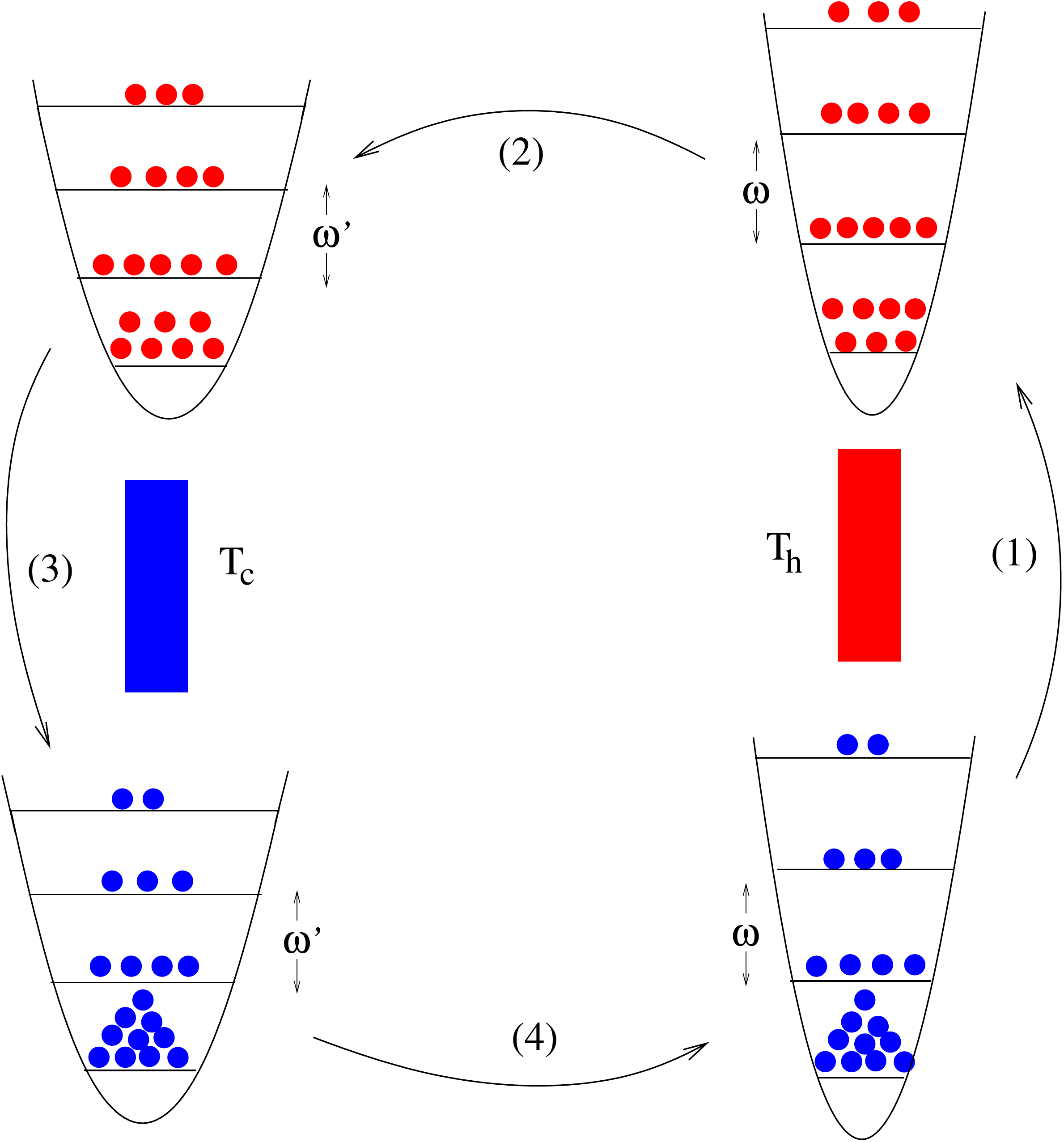}
\caption{(color online) Pictorial representation of a quantum Otto cycle. The working 
medium of this cycle is a harmonic oscillator. Stage 1 and  Stage 3 are thermalization 
processes, in which the system exchanges heat with the bath. Stages 2 and 4 correspond 
to adiabatic processes where the frequency of the oscillator changes from
 $\omega$ to $\omega'$ and back, by doing a certain amount of work.}
\label{cycle}
\end{center}
\end{figure}
\begin{itemize}
\item[] {\bf Stage 1:} In this stage, the system represented by the density matrix 
$\rho_c'$ (defined in Stage 4)  and the Hamiltonian $H$, is \textit{attached} to a 
hot bath at temperature $T_h$.  During the process, the Hamiltonian is kept fixed. 
At the end of this stage, the system reaches equilibrium with the bath. Therefore, 
the final state  is given as $\rho_h=\exp(-\beta_h H)/{\rm Tr}[\exp(-\beta_h H)]$, 
where  $\beta_h=1/k_B T_h$,  with $k_B$ being the Boltzmann constant. Hence the 
amount of heat absorbed by the system from the hot bath is $Q_h={\rm Tr}[H(\rho_h-\rho_c')]$.
\item[] {\bf Stage 2:} The system is \textit{decoupled} from the bath and the 
Hamiltonian is changed from $H$ to $H'$ slowly enough so that the quantum adiabatic
 theorem holds. Since there is no heat exchange between the system and the bath, 
the change in mean energy is equal to the work. The work done in this process is 
$w_1={\rm Tr}[(\rho_h H-\rho_h' H')]$, where $\rho_h'=U_I\rho_h U_I^{\dagger}$ 
and $U_I$ is the unitary associated with the adiabatic process, defined as 
$U_I={\cal T}\exp{[-(i/\hbar)\int_0^\tau H(t)dt]}$. Here ${\cal T}$ is the time 
ordering operator, $H(0)=H$ and $H(\tau)=H'$.
\item[] {\bf Stage 3:} The system is \textit{attached} to the cold bath at inverse 
temperature $\beta_c=1/k_B T_c$. The system reaches equilibrium with the cold bath at the 
end of the process and the state of the system becomes $\rho_c=\exp(-\beta_c H')/{\rm Tr}[\exp(-\beta_c H')]$. 
Therefore the heat rejected to the cold bath is given as $Q_c={\rm Tr}[H'(\rho_c-\rho_h')]$.
\item[] {\bf Stage 4:} The system is \textit{detached} from the cold bath 
and the Hamiltonian is slowly varied from $H'$ to $H$. The work done in this process
 is equal to the change in the mean energy, which is given as $w_2={\rm Tr}[(\rho_c H'-\rho_c' H)]$, 
where $\rho_c'$ is the density matrix at the end of the adiabatic process, defined as  
$\rho_c'=U_{II}\rho_c U_{II}^{\dagger}$ and $U_{II}$ is  given by
 $U_{II}={\cal T}\exp{[-(i/\hbar)\int_0^\tau H(t)dt]}$, so that $H(0)=H'$ and $H(\tau)=H$. 
Finally, the cycle is completed  by attaching the system with the hot bath.
\end{itemize}
The net work done by the system is $W=w_1+w_2=Q_h+Q_c$ and the efficiency is defined as
 $\eta=W/Q_h$.

In Ref. \cite{Johal2008, Armen2010},  a two-staged cycle is considered with  two n-level systems.
There, the coupling between two n-level systems is only for a short time to perform 
SWAP like operations. This cycle is equivalent to a cycle with single n-level system undergoing  a four-staged Otto cycle.
This type of cycle is  considered  in Section  \ref{sec2A}. 
Our  cycle is a four staged cycle, where the working medium consists of two oscillators (or spins)  
coupled to each other throughout the cycle. Both the subsystems
are connected to one and the same bath at a time.  Our primary aim is to find  
the effect of this coupling in the performance of the engine or refrigerator.
We analyze the performance of the system using internal parameters such as bare mode frequencies and coupling strengths.
 Later in this paper, we consider mean energy preserving co-ordinate transformations.
We refer the subsystems after the co-ordinate transformations as 'independent subsystems' since they appear to be independent 
but they may not represent the actual subsystems.
Before discussing the coupled systems, we briefly review the case of a single system as a heat engine.
\subsection{Single system as a heat engine}\label{sec2A}
Otto cycle with a single harmonic oscillator (or a spin system) constituting the working 
medium of the engine is studied in different works 
\cite{Thomas'2011,Wang'2015,Kieu'2004,Quan'2007,Rezek'2006,Chaturvedi'2013}. Here we briefly review this. 
Consider a harmonic oscillator with Hamiltonian
\begin{equation}
H^{\rm os}= \frac{p^2}{2m}+\frac{m \Omega^2}{2}x^2=\left(c^{\dagger} c+\frac{1}{2}\right)\Omega,
\label{single_os_Ham}
\end{equation}
where $m$ is the mass and $\Omega$ is the frequency of the oscillator, and
 $c^{\dagger}=x \sqrt{m\Omega/2}-ip/\sqrt{2 m\Omega}$ and $c=x \sqrt{m\Omega/2}+ip/\sqrt{2 m\Omega}$ 
are the creation and the annihilation operators respectively. We set $\hbar$ and $k_B$ to unity. 
The cycle is constructed such that in Stage 2, the frequency is changed from $ \Omega=\omega$ to 
$\Omega=\omega'$  and in Stage 4, $ \omega' $ is changed to $\omega$. In Stages 1 and 3, 
thermalization occurs with the respective heat baths as discussed above. 
The mean population of the thermal state of a harmonic oscillator with frequency $\Omega$ and inverse temperature 
$\beta$ is $\langle c^{\dagger}c\rangle=\langle n \rangle=1/(\exp{\beta\Omega}-1)$. We also 
assume that the adiabatic processes are slow enough ($\tau\rightarrow\infty$ in Stages 2 and 4), 
so that  coherence is not created between the eigenstates of the final Hamiltonian. Therefore, 
the mean population in the initial and the final states of the adiabatic process are same. Under these 
assumptions, the heat absorbed from hot reservoir is given by
\begin{eqnarray}
Q_h ^{\rm os}&=&{\rm Tr}\left[H^{\rm os}(\rho_h-\rho_c')\right]\nonumber\\
&=&\frac{\omega}{2}\left({\rm coth}\left[\frac{\beta_h\omega}{2}\right]-{\rm \coth}\left[\frac{\beta_c\omega'}{2}\right]\right),
\label{Q1_single_HO}
\end{eqnarray}
where $H^{\rm os}$ is obtained by substituting $\Omega=\omega$  in Eq. (\ref{single_os_Ham}). Here $\rho_c'$ and $\rho_h$ are respectively the 
initial and the final density matrices in Stage 1. Similarly, the heat rejected to the cold reservoir is
\begin{eqnarray}
Q_c^{\rm os}&=&{\rm Tr}\left[H'^{\rm os}(\rho_c-\rho_h')\right]\nonumber\\
&=&-\frac{\omega'}{2}\left({\rm coth}\left[\frac{\beta_h\omega}{2}\right]-{\rm \coth}\left[\frac{\beta_c\omega'}{2}\right]\right),
\label{Q2_single_HO}
\end{eqnarray}
where $H'^{\rm os}$ is calculated by substituting $\Omega=\omega'$ in Eq. 
(\ref{single_os_Ham}) and $\rho_h'$ and $\rho_c$ are the initial and final density matrices respectively
for the thermalization process described in Stage 3.
The net work done by the system  is given as $W^{\rm os}=Q_h^{\rm os}+Q_c^{\rm os}$:
\begin{equation}
W^{\rm os}= \frac{(\omega-\omega')}{2}\left({\rm coth}\left[\frac{\beta_h\omega}{2}\right]-{\rm \coth}\left[\frac{\beta_c\omega'}{2}\right]\right).
\label{work-oscillator-single}
\end{equation}
The efficiency of the system is  then given as
\begin{equation}
\eta^{\rm os}\equiv\frac{W^{\rm os}}{Q_h ^{\rm os}}=1-\frac{\omega'}{\omega}.
\label{single_osci_eff}
\end{equation}
The condition for the system to work as an engine is $W^{\rm os}\geq0$ (with $Q_h ^{\rm os}\geq0$) is satisfied when $\beta_h\omega\leq \beta_c\omega'$ and $\omega>\omega'$,
so that we have $\eta^{\rm os}\leq 1-\beta_h/\beta_c=\eta_c$, where $\eta_c$ is the Carnot efficiency. The work output is zero when the system operates at Carnot value.

Now, consider a single spin-$1/2$ system, placed under a magnetic
 field $B_z$ applied along the z-direction. To get a  similar
 form like oscillator, we need to add a term of the 
 form $\mu B_z I_2$, where $I_2$ is the identity matrix and $\mu$ is a constant.
 Adding a constant term ($\mu B_z$) with each energy eigenvalue
 does not alter the characteristics of the engine. 
 Thus we can write the corresponding Hamiltonian as,
\begin{equation}
 H^{\rm sp}= \mu B_z (S_z+I)=\left(S^+S^-+\frac{1}{2}\right)\Omega,
 \label{single_spin_H}
\end{equation}
where $S_z=\sigma_z/2$, $\Omega=\mu B_z$ and $S^+$ and $S^-$ are raising 
and lowering operators respectively. 
This Hamiltonian has a similar structure as the oscillator Hamiltonian
given in Eq. (\ref{single_os_Ham}). Now consider a cycle constructed 
such that frequency $\Omega$ varies from $\omega$ to $\omega'$ in Stage 2
 and returns to the initial value ($\omega'\rightarrow \omega$) in Stage 4. 
The heat absorbed from the hot reservoir is 
\begin{eqnarray}
Q_h ^{\rm sp}&=&{\rm Tr}\left[H^{\rm sp}(\rho_h-\rho_c')\right]\nonumber\\
&=& \frac{\omega}{2}\left({\tanh}\left[\frac{\beta_c\omega'}{2}\right]-\tanh\left[\frac{\beta_h\omega}{2}\right]\right),
\label{Q1_single_spin}
\end{eqnarray}
where $H^{\rm sp}$ is obtained by substituting $\Omega=\omega$ in Eq. (\ref{single_spin_H}). 
We also have  $ \rho_c'=\rho_c$ and 
$\rho_h'=\rho_h$, since $[U_{II},\rho_c]=0$ in Stage 4 and  $[U_{I},\rho_h]=0$ in Stage 2. 
Similar as above, the net work done by the system is given by
\begin{equation}
W^{\rm sp}= \frac{(\omega-\omega')}{2}\left({\tanh}\left[\frac{\beta_c\omega'}{2}\right]-\tanh\left[\frac{\beta_h\omega}{2}\right]\right).
\label{work-spin-single}
\end{equation}
So we can calculate the  efficiency of system as
\begin{equation}
\eta^{\rm sp}=\frac{W^{\rm sp}}{Q_h ^{\rm sp}}=1-\frac{\omega'}{\omega}.
\label{single_spin_eff}
\end{equation}
Even though the dimensionality of spin and harmonic oscillators are different, we
 kept the same energy level spacings in both the cases. Hence, both the cycles have
 the same efficiencies as shown in Eqs. (\ref{single_osci_eff}) and (\ref{single_spin_eff}).
 From Eqs. (\ref{work-oscillator-single}) and (\ref{work-spin-single}), we have
 $W^{\rm os}\ge W^{\rm sp}$.
 This inequality is true, because, for  positive real values of $x$ and $y$ 
 ($x\equiv\beta_h\omega/2<y\equiv\beta_c\omega'/2$), 
we have $(\coth{[x]}-\coth{[y]})\geq (\tanh{[y]}-\tanh{[x]})$. 

Consider two single systems (oscillators or spins), which are uncoupled and undergoing
 the cycle as discussed above.
Then the  work done by uncoupled oscillators is $2 W^{\rm os}$ is greater than the work 
done  by the spins $2 W^{\rm sp}$. But the efficiency of the 
uncoupled oscillators is equal to that of uncoupled spins, 
$\eta^{\rm os}=\eta^{\rm sp}=1-\omega'/\omega$. Now, our interest is to
 compare the performances of spins and oscillators when the analogous type of coupling is introduced.
\section{Performance of coupled  system}\label{sec3}
In this section, we study the effect of coupling in the performance of joint
 systems as the working media of Otto cycle. First, we consider two identical 
oscillators coupled via positions and momenta and then we consider spins coupled 
though Heisenberg XY model.
\subsection{Coupled  oscillators}\label{sec3A}
Consider two oscillators (labeled as 1 and 2) with same mass and frequency, and 
consider that they are coupled through their positions and momenta. The total 
Hamiltonian of the composite system reads \cite{Estes1968,Zoubi'2000,Levy2014}:
\begin{eqnarray}
H^{\rm os}&=& \frac{p_1^2}{2m}+\frac{p_2^2}{2m}+\frac{m \Omega^2}{2}x_1^2+\frac{m \Omega^2}{2}x_2^2 \nonumber \\ 
&&+2\left( \frac{m \Omega}{2}\lambda_xx_1x_2+\frac{1}{2 m\Omega}\lambda_pp_1p_2\right),
\end{eqnarray}
where $\lambda_x$ and $\lambda_p$ are the coupling strengths with  same units 
as that of $\Omega$. We can write this Hamiltonian
 in terms of ladder operators ($c_i$ and  $c^{\dagger}_i$, where $i=1,2$) as,
\begin{eqnarray}
H^{\rm os}&=&\left(c^{\dagger}_1 c_1+c^{\dagger}_2 c_2+1\right)\Omega
+\frac{(\lambda_x+\lambda_p)}{2}(c_1^{\dagger}c_2+c_1c_2^{\dagger})\nonumber \\
&&~~~~~~~~~~~~~~~~+\frac{(\lambda_x-\lambda_p)}{2}(c_1c_2+c_1^{\dagger}c_2^{\dagger}).
\label{coupled_osci_gen}
\end{eqnarray}
where $c^{\dagger}_j=x_j \sqrt{m\Omega/2}-ip_j/\sqrt{2 m\Omega}$ ($j=1,2$).
For the quadratic coupling given in Eq. (\ref{coupled_osci_gen}), let us consider the following co-ordinate transformation, 
\begin{eqnarray}
x_A=\frac{x_1+x_2}{\sqrt{2}},&\;\;\;&x_B=\frac{x_1-x_2}{\sqrt{2}};\\
p_A=\frac{p_1+p_2}{\sqrt{2}},&\;\;\;&p_B=\frac{p_1-p_2}{\sqrt{2}}.
\label{transformation}
\end{eqnarray}
The Hamiltonian in terms of new coordinates reads,
\begin{eqnarray}
H^{\rm os}&=&\frac{p_A^2}{2M_A}+\frac{M_A \Omega_A^2}{2}x_A^2+ \frac{p_B^2}{2M_B}+\frac{M_B \Omega_B^2}{2}x_B^2\\
&=& \left(c^{\dagger}_A c_A+\frac{1}{2}\right)\Omega_A+\left(c^{\dagger}_B c_B+\frac{1}{2}\right)\Omega_B,
\end{eqnarray}
where $c_k^{\dagger}$ and $c_k$, where $k={A,B}$, 
 are the creation and annihilation operators for the oscillators $A$ and $B$.
Here $\Omega_A$ and $\Omega_B$ are eigenmode frequencies and $M_A$ and $M_B$ are the effective masses in the
 new co-ordinate frame and the explicit expressions are given as,
 \begin{eqnarray}
M_{A/B}&=&\frac{m\Omega}{(\Omega\pm\lambda_p)},\\
\Omega_{A/B}&=&\sqrt{(\Omega\pm\lambda_p)(\Omega\pm\lambda_x)}
\label{frequency_expressions}.
\end{eqnarray}
Note that, in the new frame, the modes ($A$ and $B$) are uncoupled. Now consider  the above mentioned Otto
 cycle in which $\Omega$ is changed from its initial value $\omega$ to $\omega'$
 in the first adiabatic process. Correspondingly, the eigenfrequency for the oscillator $A$ changes from
 $\omega_A$ to $\omega_A'$ and similarly, frequency changes from $\omega_B$ to $\omega_B'$ in the case of oscillator $B$.
 The eigenfrequencies return to the respective initial values in the second adiabatic process. Here one can consider that the 
 working medium consists of two independent oscillators.
 Hence the total work done by the system can be
 considered as the sum of the contributions from independent oscillators.
 Note that $\omega_A$ and $\omega_B$ ( $\omega_A'$ and $\omega_B'$ ) are the effective frequencies of the subsystems after the co-ordinate
 transformation. Therefore they are functions of actual frequencies $\omega$ ($\omega'$) and coupling strengths. Hence
  $\omega_A$ and $\omega_B$ ( $\omega_A'$ and $\omega_B'$ ) are controlled by changing the frequencies of the actual subsystems. It can be
 done by changing potential in the case of oscillators or by changing the external  magnetic field
 for the spins (see Section \ref{sec3B}). We assume that there is no cross over of energy levels of the total Hamiltonian during the adiabatic process.
The density matrix may change during the adiabatic process. But the process is  slow enough such that 
the populations at the instantaneous eigenstates of the Hamiltonian remain same (quantum adiabatic theorem).
As discussed in Section \ref{sec2A}, the total amount of heat absorbed by the system from hot reservoir is given by
\begin{eqnarray}
Q&=&\frac{\omega_A}{2}\left({\rm coth}\left[\frac{\beta_h\omega_A}{2}\right]
-{\rm \coth}\left[\frac{\beta_c\omega_A'}{2}\right]\right)\nonumber \\
&+&\frac{\omega_B}{2}\left({\rm coth}\left[\frac{\beta_h\omega_B}{2}\right]
-{\rm \coth}\left[\frac{\beta_c\omega_B'}{2}\right]\right).
\label{heat}
\end{eqnarray}
The first term denotes the heat absorbed by the system $A$ ($Q_A$) and the second
 term represents the heat absorbed by the system $B$  ($Q_B$). 
Similarly, the total work is the sum of the work done by the independent systems, $W=W_A+W_B$. Thus here
\begin{eqnarray}
W&&=\frac{(\omega_A-\omega_A')}{2}\left({\rm coth}\left[\frac{\beta_h\omega_A}{2}\right]
-{\rm \coth}\left[\frac{\beta_c\omega_A'}{2}\right]\right)\nonumber \\
&+&\frac{(\omega_B-\omega_B')}{2}\left({\rm coth}\left[\frac{\beta_h\omega_B}{2}\right]
-{\rm \coth}\left[\frac{\beta_c\omega_B'}{2}\right]\right).
\label{work}
\end{eqnarray}
The efficiency of the individual system is given as $\eta_k=1-\omega_k'/\omega_k$, where $k=\{A,B\}$.
But the actual efficiency of the coupled system is defined as the ratio of total 
work over the total heat absorbed by the system. So we can write
\begin{equation}
\eta=\frac{W_A+W_B}{Q_A+Q_B}= \frac{\eta_AQ_A+\eta_BQ_B}{Q_A+Q_B}.
\label{eta_gen}
\end{equation}
When both the systems are working in engine mode (i.e., $Q_A>0$ and $Q_B>0$),
we can write the above equation as
\begin{equation}
\eta=\eta_A \alpha+\eta_B (1-\alpha)
\label{global_eta_with_A_B}
\end{equation}
where $\alpha=Q_A/(Q_A+Q_B)\leq1$. Therefore we can write
\begin{equation}
{\rm min}\{\eta_A,\eta_B\}\leq\eta\leq{\rm max}\{\eta_A,\eta_B\}.
\label{eta_inequalities}
\end{equation}
Therefore, when both  independent oscillators work as an engine, 
the actual efficiency of the engine is bounded  above and below by  the 
efficiencies of the independent oscillators. For certain parameter values (see Section \ref{sec4A}),
 one of the independent oscillators can work as refrigerator. In that case, the efficiency of the system is upper bounded by the efficiency of the other independent subsystem
 working as the engine.
\subsubsection{Generalization:}
\begin{figure}[h!]
\begin{center}
\vspace{0.2cm}
\includegraphics[width=4cm]{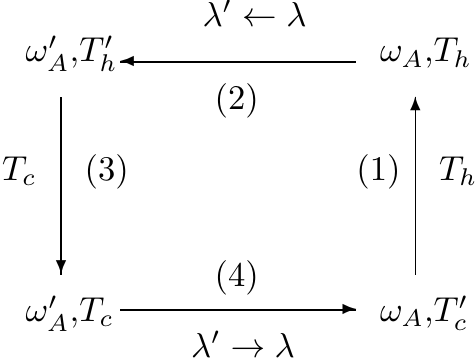}
\caption{The coupling parameter $\lambda$ is changed during the adiabatic process of quantum Otto cycle.}
\hfill
\label{pic2}
\end{center}
\end{figure}

Now consider an Otto cycle, where the global parameter  $\lambda$ ($\lambda_x$ or $\lambda_p$ or both) is externally 
controlled and varied in the adiabatic branches, keeping $\Omega$ fixed (see Fig. \ref{pic2}). 
Even in this case, using the following analysis, one can show the existence of the non-trivial bounds for the efficiency.
In this paper, we consider those quadratic couplings  for which we can  write the total
Hamiltonian as the sum of the Hamiltonians of two independent subsystems under some co-ordinate transformation.
Let us consider the Hamiltonian $H=H_A \oplus H_B$ at the end of Stage 1 and 
before the first adiabatic process (Stage 2) and 
$H'=H_A' \oplus H_B'$ be the Hamiltonian at the end of the first adiabatic process.
Before the starting of the adiabatic process,
the system is in a thermal state and hence we can write the initial state as $\rho=\rho_A\otimes\rho_B$ in the eigenbasis of $H=H_A \oplus H_B$.
Further, the adiabatic process ensures that the populations in the instantaneous eigenstates
of the Hamiltonian remain unchanged during the process.
Therefore, we can write the density matrix of  the final state
as  $\rho'=\rho_A'\otimes\rho_B'$ in the basis
of  $H_A' \oplus H_B'$.
For example, consider the coupled spin systems after the first thermalization process. Let us suppose $E_A^1$ and $E_A ^2$ are
the energy eigenvalues for the independent subsystem $A$ with the corresponding populations $p_A^1$ and $p_A^2$ ($p_A^1+ p_A^2=1$) respectively.
Similarly, $E_B^1$ and $E_B ^2$ are the energy eigenvalues for the second spin $B$ with populations $p_B^1$ and $p_B^2$ ($p_B^1+ p_B^2=1$) respectively.
So the total energy can be written as
\begin{eqnarray}
\sum_{m=1}^2 p_A^m E_A^m+\sum_{n=1}^2 p_B^n E_B^n= \sum_{k=1}^2\sum_{l=1}^2(E_A^k+E_B^l)p_A^k p_B^l
\end{eqnarray}
The terms $\{(E_A^k+E_B^l)\}$ and $\{p_A^k p_B^l\}$ in the right hand side represent 
the energy eigenvalues of the composite system and the corresponding populations respectively.
At the end of the adiabatic process, the energy eigenvalues of the composite system changes but the Hamiltonian structure allows us to write it as the sum of 
the contributions from two independent subsystems.
Therefore the  energy eigenvalues at the end of the adiabatic process are given as $\{(E_A'^k+E_B'^l)\}$. 
But the corresponding populations remains fixed because of the quantum adiabatic theorem.
Hence we can write
\begin{eqnarray}
\sum_{k=1}^2\sum_{l=1}^2(E_A'^k+E_B'^l)p_A^k p_B^l=\sum_{m=1}^2 p_A^m E_A'^m+\sum_{n=1}^2 p_B^n E_B'^n
\end{eqnarray}
Therefore the populations of the eigenstates of the independent subsystems
are also unchanged during the process.
Similar characteristics can be observed in the second adiabatic process also. Hence the total system can 
be considered as composition of two independent
subsystems in the beginning as well as at the end of each process in the cycle.
So, the total heat (or work) is the sum of the heat (or work) obtained from its independent subsystems.
Therefore, the efficiency of the system will be bounded above and below by the efficiencies of the independent systems [Eq. (\ref{eta_inequalities})].

\begin{figure}[h!]
\begin{center}
\vspace{0.2cm}
\includegraphics[width=8cm]{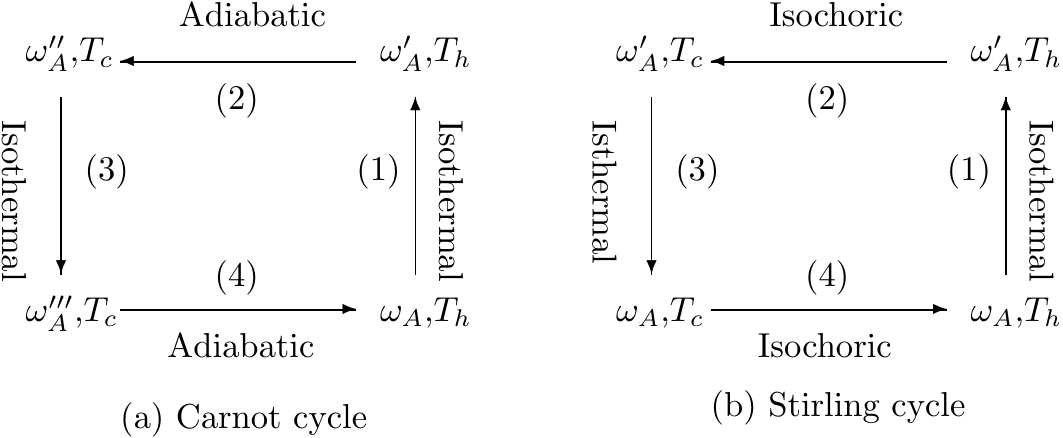}
\caption{(a) Diagram represents a Carnot cycle undergone by one of the subsystems (say A). The cycle consists of two isothermal processes and two adiabatic processes.
Here, $\omega_A$, $\omega_A'$, $\omega_A''$ and $\omega_A'''$ are the eigenmode frequencies at different stages of the cycle. (b) Diagram pictorially represents quantum Stirling cycle,
 consisting of two isochoric and two isothermal processes.}
\hfill
\label{pic3}
\end{center}
\end{figure}

An isothermal process can be simulated by an infinite number of infinitesimal
adiabatic and isochoric processes \cite{Quan'2008}.
As discussed above, in the adiabatic and isochoric processes, 
the work and the heat contributions from individual subsystems can be identified separately. 
Therefore, the isothermal process of the total system 
can be considered equivalent to the isothermal processes of two independent
subsystems taken together. To understand the generality of the bounds of the efficiency observed in Otto cycle,
we can consider other cycles such as Carnot cycle consisting of two isothermal processes and two
adiabatic processes, and Stirling cycle consisting of two isothermal processes and two
isochoric processes \cite{Quan'2007,Chaturvedi'2013,Huang2014_Stirling} (see Fig. \ref{pic3}).
Like the Otto cycle, in the Stirling cycle,  the global efficiency is bounded by the efficiencies of 
independent subsystems because at any stage of the cycle, the total
system can be decomposed into two independent subsystems.
This shows that the analysis on the bounds of the efficiency represented
 in Eq. (\ref{eta_inequalities}) is a generic one and applicable to
 some of the other cycles also. Using similar analysis, we can also show that
for a quantum Carnot cycle,  both the independent 
subsystems work at Carnot efficiency, which is same as the global efficiency, i.e., 
$\eta=\eta_A=\eta_B=1-T_c/T_h$.
\subsection{Coupled spin system}\label{sec3B}
In order to compare the performance of the quantum Otto cycle with coupled 
oscillator and that with coupled spin-1/2 system, let us now 
consider two spin-1/2 systems coupled via Heisenberg exchange interaction,
 placed in a magnetic field applied  along  the $z$-direction. The Hamiltonian in terms of spin operators
are given by
\begin{equation}
 H^{\rm sp}=\Omega(S^z_1\otimes I+ I\otimes S^z_2) +2 (J_x S^x_1S_2^x +J_y S^y_1S_2^y),
\end{equation}
where $J_x$ and $J_y$ are the interaction constants along $x$ and $y$ directions.  
This model is generally known as Heisenberg XY model.  Adding an equal
 energy with each level, we can write the Hamiltonian in terms of  raising and lowering operators
($S^+_i$ and  $S^-_i$, where $i=1,2$)  as
\begin{eqnarray}
 H^{\rm sp}&=&(S^{+}_1S_1^-+S^{+}_2S_2^-+1)\Omega\nonumber\\
 &&+\frac{(J_x+J_y)}{2}(S_1^+S_2^-+S_1^-S_2^+)\nonumber\\
 &&+\frac{(J_x-J_y)}{2}(S_1^+S_2^++S_1^-S_2^-).
 \label{coupled_spin_gen}
\end{eqnarray}
Equations (\ref{coupled_osci_gen}) and  (\ref{coupled_spin_gen}) have the similar form. 
As we have seen in the case of oscillators,
Eq. (\ref{coupled_spin_gen}) can be written as the sum of the Hamiltonians
of independent subsystems. 
Therefore, the efficiency of the system is bounded from above and below by the 
efficiencies of independent subsystems as given in Eq. (\ref{eta_inequalities}).
In the following section (Sec. \ref{sec4}), we compare
the performances of coupled spins and coupled oscillators when they undergo separately 
the quantum Otto cycles for different values of  $\lambda_x$,  $\lambda_p$, $J_x$, and $J_y$.
\subsection{Optimal work} \label{sec3C}
As we have seen in Eqs. (\ref{work-oscillator-single}) and (\ref{work-spin-single}),
work is a function of $\omega$ and $\omega'$.  The optimal work can be estimated by 
maximizing work with respect to $\omega$ and $\omega'$.
Now suppose that work is maximum  at  $\omega=\omega^*$ and $\omega'=\omega'^*$. Therefore, for two
uncoupled oscillators, maximum work  occurs when both the system work with $\omega^*$ and $\omega'^*$. 
Now consider $\Omega_A$ and $\Omega_B$  are frequencies of the independent modes $A$ and $B$ of the coupled oscillators respectively,
before the first adiabatic process and $\Omega_A'$ and $\Omega_B'$ are the frequencies of the independent modes $A$ and $B$ respectively, after the first adiabatic process.
Then, $\Omega_A$ and $\Omega_B$ ($\Omega_A'$ and $\Omega_B'$) are functions 
of $\omega$, $\lambda_x$ and $\lambda_p$ ($\omega'$, $\lambda_x$ and $\lambda_p$). 
Similar arguments can be made for coupled spins too.
Therefore, if the subsystem $A$ provides optimal work, then the work obtained from the 
subsystem $B$ may not be optimal because $\Omega_A$ and $\Omega_B$ ($\Omega_A'$ and $\Omega_B'$) are not independent. Therefore we have
\begin{equation}
 W_{\lambda}^{\rm max}\leq W_{0}^{\rm max},
 \label{optimal work}
\end{equation}
$ W_{\lambda}^{\rm max}$ and $W_{0}^{\rm max},$ are  the maximum values of work
obtained from the coupled and uncoupled systems respectively. The equality holds for the case 
where both  independent subsystems have the same frequency.

\subsubsection{Generalization:}
Now consider a cycle in which the global parameters $\lambda_x$ and $\lambda_p$ or $J_x$ and
$J_y$ are varied in the adiabatic branches instead of varying $\Omega$. Because of quadratic coupling,  we can show that the total system consists of 
two independent subsystems, throughout the cycle. Then  the total work is the 
sum of the contributions from its independent subsystems and therefore Eq. (\ref{optimal work}) holds even when we change  global parameters  to extract
work.

In Stirling cycle and in Carnot cycle, for a working medium with quadratic interaction, 
we can always write the total work as the sum of the  contributions from its independent subsystems.
Therefore, using the  argument mentioned above, we can say that the Eq. (\ref{optimal work}) holds in general. 
 
 In sections \ref{sec4} and \ref{sec5}, we discuss the Otto cycle in detail by considering certain special cases.
\section{Special cases}\label{sec4}
In this section, we discuss the performance of the Otto cycle when the coupled systems with specific 
values of interaction constants constitute as the working medium. In  Section \ref{sec4A}, we take $J_x=J_y$ in 
the case of spin, which is known as
Heisenberg XX model. Analogous interaction in oscillators is achieved by setting
 $\lambda_x =\lambda_p$. Another interesting model is obtained with values $J_x=-J_y$ in spins and
$\lambda_x =-\lambda_p$ in oscillators, are discussed in Section \ref{sec4B}.
\subsection{XX model}\label{sec4A}
Let us  consider the following case:  $\lambda_x=J_x =\lambda_p=J_y=\lambda_J ({\rm say})$.
For coupled oscillators, we can write the Hamiltonian in Eq. (\ref{coupled_osci_gen}) in terms of  ladder operators as 
\begin{eqnarray}
 H^{\rm os}=\left(c^{\dagger}_1 c_1+c^{\dagger}_2 c_2+1\right)\Omega 
+ \lambda_J (c_1^{\dagger}c_2+c_1c_2^{\dagger}).
\label{hamiltonian_xx_os}
\end{eqnarray}
From Eq. (\ref{frequency_expressions}), we get the frequencies of the independent modes as
$\Omega_A=\Omega+\lambda_J$ and  $\Omega_B=\Omega-\lambda_J$. Therefore, inside the 
Otto cycle where the value of  $\Omega$ is varied from $\omega$ to $\omega'$ during the first adiabatic process,  
we have $\omega_{A/B}=\omega\pm\lambda_J$ and $\omega_{A/B}'=\omega'\pm\lambda_J$. 
In the new co-ordinates, the oscillators are independent. Hence the total 
heat absorbed from the hot reservoir is the sum of the heat absorbed by the independent subsystems $A$ and $B$. 
Substituting the  values of $\omega_A$, $\omega_A'$, $\omega_B$ and $\omega_B'$   in the Eqs. (\ref{heat}) and (\ref{work}),
 we get the expressions for the heat and the work respectively. 
 The explicit expression for the work obtained from  Eq. (\ref{work}) is given as
\begin{eqnarray}
W^{\rm os}&&=C\left({\rm coth}\left[\frac{(\omega-\lambda_J)}{2T_h}\right]
-{\rm \coth}\left[\frac{(\omega'-\lambda_J)}{2T_c}\right]\right)\nonumber \\
&+&C\left({\rm coth}\left[\frac{(\omega+\lambda_J)}{2T_h}\right]
-{\rm \coth}\left[\frac{(\omega'+\lambda_J)}{2T_c}\right]\right),
\label{work_os}
\end{eqnarray}
where $C=(\omega-\omega')/2$. The efficiency of the independent subsystems are obtained as
\begin{equation}
 \eta_A=\frac{(\omega-\omega')}{(\omega+\lambda_J)}\;\;\; {\rm and}\;\;\; \eta_B=\frac{(\omega-\omega')}{(\omega-\lambda_J)}.
 \label{eta_A_and_B}
\end{equation}
So we have $\eta_B>\eta_A$. Interestingly, the upper bound  $\eta_B$  is analogous to the
 upper bound of the efficiency obtained for the coupled spins with
 isotropic Heisenberg Hamiltonian \cite{Thomas'2011}.
Now we calculate the global efficiency  as the ratio of the total work ($W_A+W_B$) by the total heat ($Q_A+Q_B$). 
We get the global efficiency  by substituting the  values of $\omega_A$, $\omega_A'$, $\omega_B$  and $\omega_B'$  in the Eq. (\ref{eta_gen}).
Now we expand this efficiency for small coupling constant $\lambda_J$ up to the 
third order, and we get:
 \begin{eqnarray}
 \eta^{\rm os}=1-\frac{\omega'}{\omega}&+&
 \frac{\gamma\left(T_c {\rm csch}^2\left[\frac{\omega}{2 T_h}\right]
 -T_h {\rm csch}^2\left[\frac{\omega'}{2 T_c}\right]\right)\lambda_J^2} {2\left(\coth\left[\frac{\omega}{2 T_h}\right]
 -\coth\left[\frac{\omega'}{2 T_c}\right]\right)}\nonumber\\
 &+& O[\lambda_J^4],
 \label{eff_os_exp}
\end{eqnarray}
where $\gamma=(\omega-\omega')/(T_hT_c\omega^2)$ and $\mbox{csch}(x)\equiv {\rm cosech}(x)$. 

\begin{figure}[ht]
\begin{center}
\vspace{0.2cm}
\includegraphics[width=8cm]{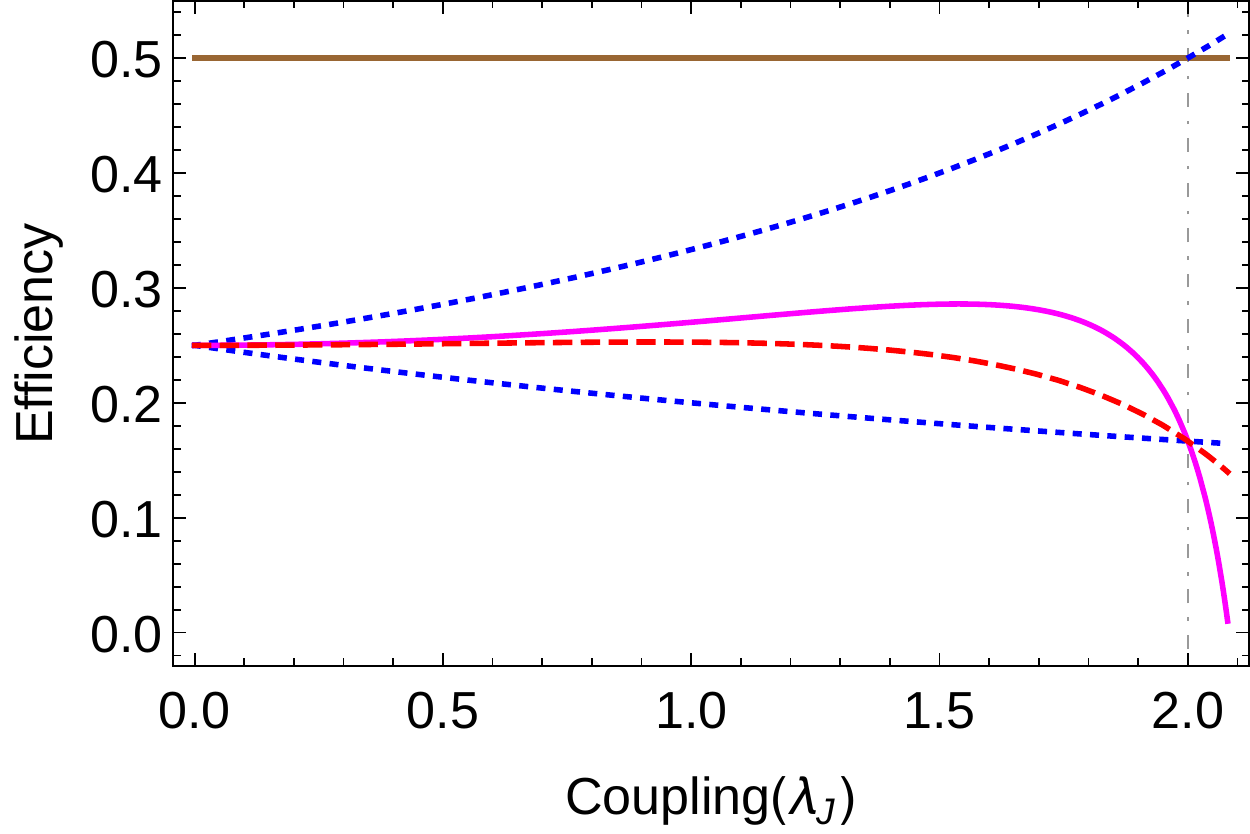}
\caption{(color online) The two dotted curves show the upper bound 
($\eta_B$) and lower bound ($\eta_A$) . The continuous curve
 represents the efficiency of the coupled oscillator. The efficiency of the coupled spin system is
denoted by the dashed curve. Carnot value is represented by the horizontal line. 
When the independent systems work in engine mode, the global efficiency of the coupled system lies inside the bounds.
The plot also shows that the global efficiency of the coupled oscillators
is higher than that of the coupled spins for small values of $\lambda_J$ 
(see Eq. (\ref{eta_os-sp})). When the upper bound reaches Carnot value, $\eta_B=1-T_c/T_h=\eta_c$
for $\lambda_J=\lambda_c$ (represented by vertical dashed-dotted line), 
then we get $\eta^{{\rm os}}=\eta^{{\rm sp}}=\eta_A$. Here we take $T_h=2$, $T_c=1$, $\omega=4$ and $\omega'=3$. 
Note that, when $\eta_B>\eta_c$, $\eta_B$ does not represent efficiency because the subsystem $B$ works as a refrigerator.}
\hfill
\label{fig1}
\end{center}
\end{figure}

We need to compare  the performance of the oscillator system with that of 
the coupled spin system as a heat engine. For that, we consider
two spin-1/2 systems coupled via Heisenberg XX Hamiltonian ($J_x=J_y$).
Representing this Hamiltonian
 in terms of ladder operators, it takes similar form of the Hamiltonian that 
we considered in the case of oscillators (see below). Therefore we write,
\begin{equation}
 H^{\rm sp}=(S^{+}_1S_1^-+S^{+}_2 S_2^-
 +1)\Omega+ \lambda_J(S_1^+S_2^-+S_1^-S_2^+).
 \label{hamiltonian_xx_sp}
\end{equation}
Coefficient $\Omega$ of $\left(c^{\dagger}_1 c_1+c^{\dagger}_2 c_2+1\right)$ in Eq. (\ref{hamiltonian_xx_os}) is characteristically same as the
coefficient $\Omega$ of $(S^{+}_1S_1^-+S^{+}_2 S_2^-
 +1)$ in Eq. (\ref{hamiltonian_xx_sp}). Same is true with the coefficient $\lambda_J$ appeared in Eqs. (\ref{hamiltonian_xx_os}) and (\ref{hamiltonian_xx_sp}). To compare the performance of coupled spins and oscillators, we can diagonalize 
the Hamiltonian for the coupled spins
so that in the new basis, spins are uncoupled. So we can write
\begin{equation}
  H^{\rm sp}=(\Omega+\lambda_J)(S^+_AS^-_A +\frac{1}{2})+(\Omega-\lambda_J)(S^+_BS_B^- +\frac{1}{2}).
\end{equation}
Therefore, the total  heat exchanged between the system and the hot bath
is the sum of contributions from spins A and B.
So we get the  heat exchanged between system $k$ ($=A$ or $B$) and the hot bath as (see Eq. (\ref{Q1_single_spin}))
\begin{equation}
 Q_k^{\rm sp}= \frac{\omega_k}{2}\left({\rm tanh}\left[\frac{\omega_k')}{2 T_c}\right]
-{\rm \tanh}\left[\frac{\omega_k}{2T_h}\right]\right).
\end{equation}
The total heat exchange between the system and the hot bath is $ Q^{\rm sp}= Q_A^{\rm sp}+ Q_B^{\rm sp}$
The total work done by the system is the sum of the contributions from 
the individual spins defined in the new basis. So we get 
\begin{eqnarray}
W^{\rm sp}&=&\eta_A  Q_A^{\rm sp}+\eta_B  Q_B^{\rm sp}\nonumber\\
W^{\rm sp}&=&C\left({\rm tanh}\left[\frac{(\omega'-\lambda_J)}{2T_c}\right]
-{\rm \tanh}\left[\frac{(\omega-\lambda_J)}{2T_h}\right]\right)\nonumber \\
&+&C\left({\rm tanh}\left[\frac{(\omega'+\lambda_J)}{2T_c}\right]
-{\rm \tanh}\left[\frac{(\omega+\lambda_J)}{2T_h}\right]\right),
 \label{work-spin}
\end{eqnarray}
with $C=(\omega-\omega')/2$. Here, $\eta_A$ and $\eta_B$ are same as that obtained in coupled oscillators given in
 Eq. (\ref{eta_A_and_B}). Therefore, the efficiency of the engine is given as
\begin{equation}
 \eta^{\rm sp}=\frac{\eta_A Q_A^{\rm sp}+\eta_B Q_B^{\rm sp}}{ Q_A^{\rm sp}+Q_B^{\rm sp}}.
\end{equation}
We can expand this efficiency for small values of $\lambda_J$ as
 \begin{eqnarray}
 \eta^{\rm sp}=1-\frac{\omega'}{\omega}&+&
 \frac{\gamma\left(T_h {\rm sech}^2\left[\frac{\omega'}{2 T_c}\right]
 -T_c {\rm sech}^2\left[\frac{\omega}{2 T_h}\right]\right)\lambda_J^2} {2\left(\tanh\left[\frac{\omega}{2 T_h}\right]
 -\tanh\left[\frac{\omega'}{2 T_c}\right]\right)}\nonumber\\
 &+& O[\lambda_J^4].
\label{eff_sp_exp}
\end{eqnarray}
The difference in the efficiencies obtained from coupled oscillators 
and coupled spins for small coupling is calculated
from Eqs. (\ref{eff_os_exp}) and  (\ref{eff_sp_exp}). When the coupling
$\lambda_J=0$, both  oscillator as well as spin system yields the same 
efficiency as discussed in Section \ref{sec2}. 
 By introducing a small coupling between the systems, we can write the difference
between the efficiencies of the oscillator model and the spin model as
 \begin{eqnarray}
 \eta^{\rm os}-\eta^{\rm sp}&=& \gamma\left(T_c {\rm csch}\left[\frac{\omega}{T_h}\right]+T_h {\rm csch}\left[\frac{\omega'}{T_c}\right]\right)\lambda_J^2\nonumber\\
 &>&0,
 \label{eta_os-sp}
 \end{eqnarray}
since  $\omega>\omega'$ and $\gamma>0$. Hence in this model, for small 
values of $\lambda_J$, the efficiency achieved by coupled oscillators is higher than
the efficiency obtained from coupled spin model. 
Now we can see the behavior of the efficiency as a function of
$\lambda_J$ (see Fig. \ref{fig1}). The characteristics
of the efficiencies in Fig. \ref{fig1} remain the  same with the choice of different values of $T_h$, $T_c$, $\omega$, and $\omega'$.
When $\omega_A'/T_c\geq \omega_A/T_h$ ($\omega_A'< \omega_A$) and $\omega_B'/T_c\geq \omega_B/T_h$ ($\omega_B'< \omega_B$),
both the independent systems work as engines. 
It is interesting to note that when $\lambda_J= (\omega' T_h-\omega T_c)/(T_h-T_c)=\lambda_c\;\; ({\rm say})$, the upper bound of the efficiency,
which is the efficiency of oscillator $B$, attains the Carnot value ($\omega_B'/T_c= \omega_B/T_h$)
with zero work output. At this point, the  global efficiency of the system 
is equal to the efficiency of oscillator $A$, $\eta^{\rm os}=\eta^{\rm sp}=\eta_A$.
When $\lambda_J>\lambda_c$, oscillator $B$ works as  refrigerator, which in turn
reduces the efficiency of the engine. Hence the efficiency of the total system 
lies below the efficiency of oscillator $A$. Now we can compare the performance 
in terms of work
given in Eqs. (\ref{work_os}) and (\ref{work-spin}). The terms on the right-hand 
side of these equations are positive when both the independent systems work in the engine mode.
In that case, using the analysis made in Sec. \ref{sec2} for uncoupled systems, 
we can show that $W^{\rm os}>W^{\rm sp}$.
\subsection{Optimal work and correlations}\label{optimum_work_concurrence}
\begin{figure}[h]
\begin{center}
\vspace{0.2cm}
\includegraphics[width=10cm]{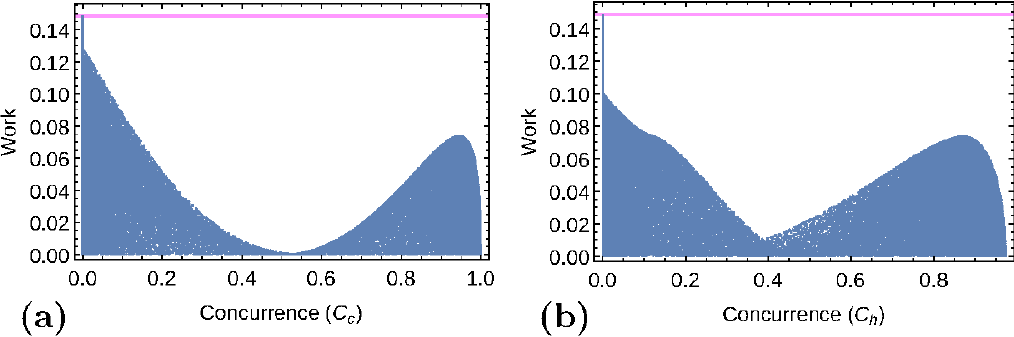}
\caption{The figure shows the behavior of work versus concurrence of the coupled spin systems (XX model).
In (a), the total work is plotted
versus concurrence at the end of stage 3 ($C_c$) and in (b) 
total work versus concurrence at the end of stage 1 ($C_h$) is plotted. 
The horizontal line represents the maximum work for an uncoupled system which is only obtained for a zero concurrence.
The temperature of the hot and the cold baths are fixed at,
$T_h=2$ and $T_c=1$ respectively. The
parameters $\omega$, $\omega'$ and $\lambda_J=J_x=J_y$ are randomly chosen
from  $[0, 10]$ such that the system should work as an engine. 
Each point in the plot refers to a given values of $\omega$, $\omega'$ and $\lambda_J$. 
There are more than $10^5$ points in  the plot.
}
\hfill
\label{concurrence}
\end{center}
\end{figure}

n the thermodynamic cycle, the origin of any correlation between the actual subsystems is due to the coupling.
When the system thermalizes at the end of stage 1 and stage 3, 
the state of the system becomes a product state irrespective of the initial state, if there is no coupling present in the system.
The proof that the optimum work extractable from
a coupled system is upper bounded by the optimum work obtained from the uncoupled system (Eq. (\ref{optimal work})) implies that the presence of quantum correlations do not have any advantage in optimal work extraction.
To illustrate this fact (see also Section \ref{sec3C}) with an example, we analyze the behavior of work versus concurrence for 
the coupled spin systems. 
The concurrence is a measure of entanglement of an arbitrary two qubit state \cite{Wootters1997,Wootters1998,Vedral2001}. 
The concurrence has one to one correspondence with the entanglement of formation. The concurrence of a state $\rho$ is defined as 
\begin{equation}
 C={\rm Max}\{0, \sqrt{\lambda_1}-\sqrt{\lambda_2}-\sqrt{\lambda_3}-\sqrt{\lambda_4}\}
\end{equation}
where $\lambda_1$, $\lambda_2$, $\lambda_3$ and $\lambda_4$ are the eigenvalues  of the matrix $R$ written in descending order.
The Matrix $R$ is defined as
$R= \rho (\sigma_y\otimes\sigma_y)\rho^* (\sigma_y\otimes\sigma_y)$ where $\sigma_y$ is the Pauli matrix.
We estimate the concurrences for the thermal
states at the end of Stage 1  and Stage 3 denoted as $C_{h}$ and $C_{c}$ respectively \cite{Zhang2007}.
For the numerical analysis shown in Fig. \ref{concurrence}, we fix the temperatures of the
hot and the cold bath, then we randomly choose the values of $\omega$, $\omega'$ and $\lambda_J=J_x=J_y$
such that system should work as an engine.
The optimum work is obtained only for zero concurrence and hence 
the numerical analysis shows that our theoretically obtained bound for the work holds.
In the Fig. \ref{concurrence}, the  work decreases and then increases with the concurrence. 
This is due to the fact that, for the coupled systems, the engine can
work in two regimes, $\omega>\omega'$ as well as $\omega'>\omega$ as discussed in \cite{Thomas'2011}.
The maximum work at given values of $C_{h}$ and $C_{c}$ is another interesting direction to study.
\subsection{XY model}\label{sec4B}
Here we  consider the  case  $ \lambda_x=J_x =-\lambda_p=-J_y=\lambda_J$ (say). 
The Hamiltonian corresponding to the coupled oscillators is now written as
\begin{equation}
 H^{\rm os}=\left(c^{\dagger}_1 c_1+c^{\dagger}_2 c_2
 +1\right)\Omega+ \lambda_J(c_1^{\dagger}c_2^{\dagger}+c_1c_2).
\end{equation}
In the new co-ordinate system (described by Eq. (\ref{frequency_expressions})),
 both the independent oscillators have the same frequency,
which is given by $\Omega_A=\Omega_B=\sqrt{(\Omega^2-\lambda_J^2)}$. 
Therefore, in the cycle, we have $\omega_A=\omega_B=\sqrt{(\omega^2-\lambda_J^2)}$ and
$\omega_A'=\omega_B'=\sqrt{(\omega'^2-\lambda_J^2)}$. Hence, we get $\eta_A=\eta_B=1-\omega_A'/\omega_A$ and
from Eq. (\ref{global_eta_with_A_B}), we obtain
\begin{equation}
 \eta^{\rm os}=\eta_A=1-\sqrt{\frac{\omega'^2-\lambda_J^2}{\omega^2-\lambda_J^2}}.
\label{eta_os_xy}
\end{equation}
On the other hand, in the spin case, the  analogous Hamiltonian  is an 
example of  Heisenberg XY model. In this case, the interaction Hamiltonian has the following
form $H_{int}=J(S_1^xS_2^x-S_1^yS_2^y)$, where $J$ is the interaction constant. This model is well studied 
as quantum Otto cycle \cite{Kosloff'2002,Kosloff'2003}. In terms of raising and lowering operators, we can
write the spin Hamiltonian as
\begin{eqnarray}
 H^{\rm sp}&=&(S^{+}_1S_1^-+S^{+}_2 S_2^-
 +1)\Omega+ \lambda_J(S_1^+S_2^++S_1^-S_2^-)\nonumber\\
 &=&\Omega_A(S^+_AS^-_A +\frac{1}{2})+\Omega_B(S^+_BS^-_B +\frac{1}{2}),
 \label{xy_spin_hamiltonian}
\end{eqnarray}
where $\Omega_A=\Omega_B=\sqrt{\Omega^2+\lambda_J^2}$.
So we have two independent spins with the same frequency. So one would expect the  
efficiency of this system is similar to that of a single system (or two uncoupled systems) given in Eq. (\ref{single_spin_eff}). 
This is true only when Stage 2 and Stage 4 are done slow enough.
Because, in this case, the eigenvectors of the Hamiltonian are functions of $\Omega$ and $\lambda_J$. 
Hence, when the system works as  Otto cycle, by changing the magnetic field associated with the system,
internal friction appears to be depending upon the rate at which magnetic field is changed \cite{Kosloff'2002,Kosloff'2003}. 
This is due to the non-commutativity of the Hamiltonian 
at different instances during the driving. Here the adiabatic processes are done 
slow enough so that quantum adiabatic theorem holds. Since the energy level spacings 
for both
the independent subsystems are equal ($\Omega_A=\Omega_B$)  in Eq. (\ref{xy_spin_hamiltonian}), 
these subsystems undergo identical cycles with the same efficiency. 
Therefore, the global efficiency is also equal to the 
efficiency of the subsystem, which is given by
\begin{equation}
 \eta^{\rm sp}=1-\sqrt{\frac{\omega'^2+\lambda_J^2}{\omega^2+\lambda^2_J}}.
 \label{eta_sp_xy}
\end{equation}
Now comparing Eqs. (\ref{eta_os_xy}) and (\ref{eta_sp_xy}), we get $\eta^{\rm os}\geq\eta^{\rm sp}$, 
for a range of parameter such that the systems work as engines (where $|\lambda_J|$ may be also large).
For a small coupling, we can expand the
efficiencies of the coupled oscillators and coupled spin systems as
\begin{equation}
 \eta_{\rm os}=1-\frac{\omega'}{\omega}+\frac{(\omega^2-\omega'^2)\lambda_J^2}{2\omega^3\omega'}+O[\lambda_J^4]
 \label{xy_osc_eta_exp}
\end{equation}
\begin{equation}
 \eta_{\rm sp}=1-\frac{\omega'}{\omega}-\frac{(\omega^2-\omega'^2)\lambda_J^2}{2\omega^3\omega'}+O[\lambda_J^4]
 \label{xy_spin_eta_exp}
 \end{equation}
Equation (\ref{xy_osc_eta_exp}) shows that when  the coupling is introduced, the efficiency of the coupled 
oscillators goes above the efficiency of the uncoupled model $\eta_{\rm uc}=1-\omega'/\omega$ [Eqs. (\ref{single_osci_eff}) and (\ref{single_spin_eff})], 
while, according to Eq. (\ref{xy_spin_eta_exp}), the efficiency of the
spin system lies below that of the uncoupled model $\eta_{\rm os}>\eta_{\rm uc}>\eta_{\rm sp}$ (see  Fig.\ref{zeta_xy_fig}).
Interestingly, Eqs. (\ref{xy_osc_eta_exp}) and (\ref{xy_spin_eta_exp}) shows the symmetric  behavior of $\eta_{\rm os}$ and $\eta_{\rm sp}$  about $\eta_{\rm uc}$.
This behavior is obvious due to the fact that coupling reduces the frequencies of the eigenmodes of the harmonic oscillators while
it increases the energy level spacings of the independent spin systems.
\section{Performance as a refrigerator}\label{sec5}
The coupled spins and coupled oscillators can also work as a refrigerator 
\cite{Kosloff2010,Kosloff2012}. In this section, we consider the case where 
both the independent subsystems
(of coupled oscillators and coupled spins separately) work as
refrigerators. The refrigeration cycle is same as the cycle described for engine 
in Section \ref{sec2}, provided refrigerators absorb heat from the
cold bath ($Q_c>0$) and  transfer it into the hot bath ($Q_h<0$). 
To transfer heat from the cold bath to the hot bath, work has to be done on the 
system and hence, we have $W=Q_h+Q_c<0$. The coefficient of performance
(COP) is defined as $\zeta=Q_c/|W|$ \cite{Chen1990,GJ2015}.

If we consider a single spin or a single oscillator (See Section \ref{sec2}) 
such that the conditions on the  parameters during the cycle are: $\omega>\omega'$ and
$\omega/T_h>\omega'/T_c$, then these systems (spin or oscillator) work as refrigerators. So we get the COP as
\begin{equation}
 \zeta^{\rm os}=\zeta^{\rm sp}=\frac{Q_c^{\rm os(\rm sp)}}{|W^{\rm os(\rm sp)}|}=\frac{\omega'}{\omega-\omega'}.
\label{single_osci_spn_zeta}
\end{equation}
Therefore, for uncoupled oscillators and spins, COPs are equal. Now, consider 
the COP for coupled systems as described in Section \ref{sec3}. Suppose $\omega_A$ and $\omega_B$ are
the frequencies of the subsystems (of  coupled oscillators or coupled spins) $A$ and $B$
respectively, before the
first adiabatic process and $\omega_A'$ and $\omega_B'$ are the frequencies of $A$ 
and $B$ at the end of the first adiabatic process. Then both the independent subsystems work as
refrigerator when $\omega_A/T_h>\omega_A'/T_c$ and $\omega_B/T_h>\omega'_B/T_c$ 
with the conditions $\omega_A>\omega_A'$ and $\omega_B>\omega_B'$.
Therefore, the local COP is $\zeta_k=\omega_k'/(\omega_k-\omega_k')$, where $k=A,B$. 
Further, the  global COP is written as
\begin{eqnarray}
 \zeta&=&\frac{\zeta_A |W_A|+\zeta_B |W_B|}{|W_A+W_B|}\nonumber\\
 &=&\alpha'\zeta_A+ (1-\alpha')\zeta_B,\nonumber
\end{eqnarray}
where $\alpha'=|W_A|/(|W_A+W_B|)<1$. Since $W_A<0$ and $W_B<0$, we have $|W_A|+|W_B|=|W_A+W_B|$.
Hence we can write
\begin{equation}
{\rm min}\{\zeta_A,\zeta_B\}\leq\zeta\leq{\rm max}\{\zeta_A,\zeta_B\}.
\label{zeta_inequalities}
\end{equation}
The global COP is bounded by COPs of the subsystems when both the  subsystems work as refrigerators.
Now our task is to understand the behavior of COPs for special cases as we have done for
heat engine in Section \ref{sec4}.
\subsection{XX model}\label{sec5A}
\begin{figure}[ht]
\begin{center}
\vspace{0.2cm}
\includegraphics[width=8cm]{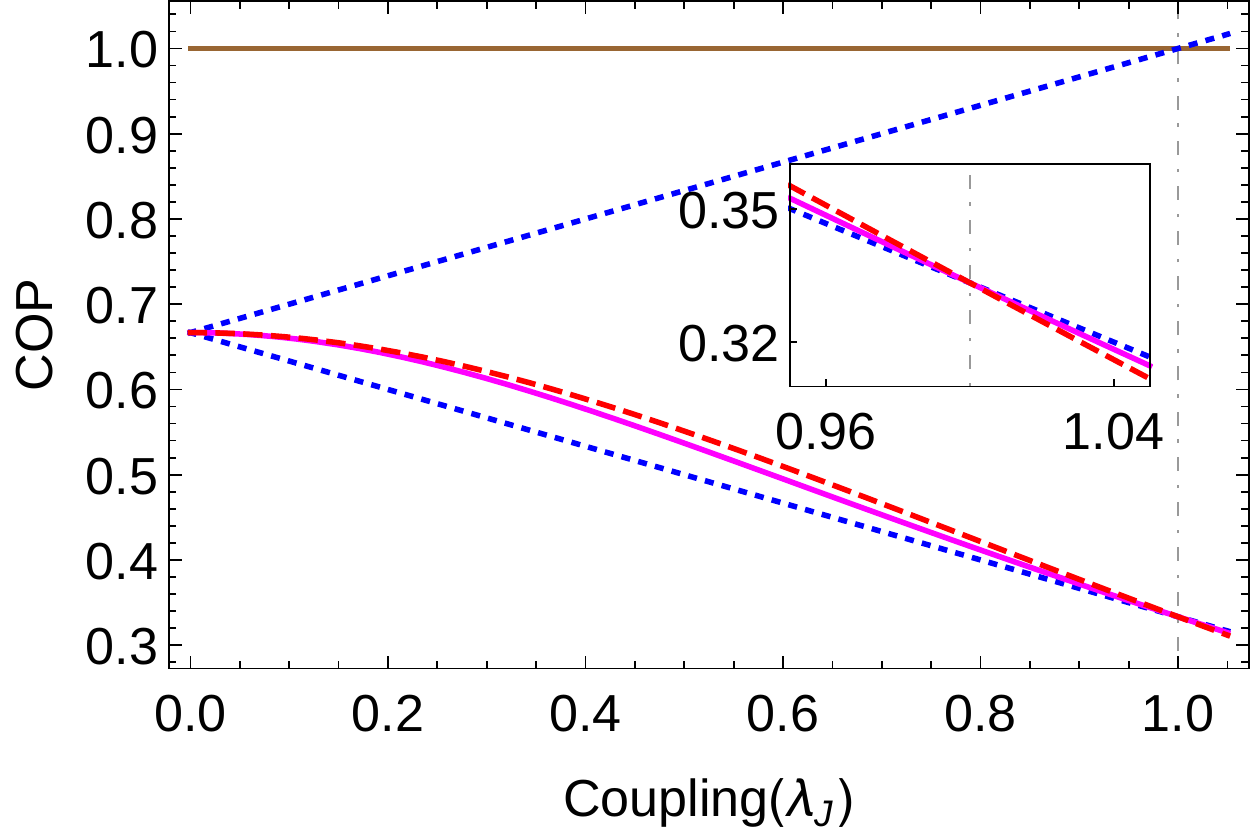}
\caption{(color online) The upper bound ($\zeta_A$) and the lower bound ($\zeta_B$) 
are shown with the dotted curves. The continuous curve
 represents the COP of the coupled oscillators while the COP of the coupled spin system is
denoted by the dashed curve. The horizontal line represents the Carnot value for the refrigerator. 
When the independent subsystems work in refrigerator mode, the global COP of the
 coupled system is bounded by $\zeta_A$ and $\zeta_B$.
The plot also shows that the global COP of the coupled spins
is higher than that of the coupled oscillators for small values of $\lambda_J$ as
 seen in Eq. (\ref{zeta_os-sp}). When the upper bound achieves Carnot value, $\zeta_A=T_c/(T_h-T_c)=\zeta_c$
for $\lambda_J=\lambda_c'$ (shown by vertical dashed-dotted line), thus we 
get $\zeta^{{\rm os}}=\zeta^{{\rm sp}}=\zeta_B$. The inset shows the enlarged region 
near $\lambda_J=\lambda_c'$. Here we take $T_h=2$, $T_c=1$, $\omega=5$ and $\omega'=2$.
Note that, when $\zeta_A>\zeta_c$, $\zeta_A$ does not represent COP, as the subsystem $A$ works as an engine.}
\hfill
\label{fig_cop}
\end{center}
\end{figure}

Here we compare the COPs of coupled oscillators and coupled spins when the coupling in each case is of
$XX$ type. The COPs of subsystems are obtained
as $\zeta_A=(\omega'+\lambda_J)/(\omega-\omega')$
and $\zeta_B=(\omega'-\lambda_J)/(\omega-\omega')$ and hence we have $\zeta_A>\zeta_B$.
We calculate the global COP as $\zeta^{\rm os}=Q_c^{\rm os}/|W^{\rm os}|$. 
Expanding for small values of $\lambda_J$, we get 
\begin{eqnarray}
 \zeta^{\rm os}&&=\frac{\omega'}{(\omega-\omega')}
 +\frac{\left(T_h{\rm csch}^2\left[\frac{\omega'}{2T_c}\right]-T_c{\rm csch}^2\left[\frac{\omega}{2T_h}\right]\right)\lambda_J^2}
 {2\gamma'\left(\coth\left[\frac{\omega}{2T_h}\right]-\coth\left[\frac{\omega'}{2T_c}\right]\right)}\nonumber\\
 &&+O[\lambda_J^4],
 \label{zeta_os_xx}
\end{eqnarray}
where $\gamma'= T_hT_c(\omega-\omega')$. Similarly, for the spin systems, the COP
 is $\zeta^{\rm sp}=Q_c^{\rm sp}/|W^{\rm sp}|$. Further, expanding $\zeta^{\rm sp}$, we get
\begin{eqnarray}
 \zeta^{\rm sp}=\frac{\omega'}{(\omega-\omega')}
 &+&\frac{\left(T_c{\rm sech}^2\left[\frac{\omega}{2T_h}\right]-T_h{\rm sech}^2\left[\frac{\omega'}{2T_c}\right]\right)\lambda_J^2}
 {2\gamma'\left(\tanh\left[\frac{\omega}{2T_h}\right]-\tanh\left[\frac{\omega'}{2T_c}\right]\right)}\nonumber\\
 &+&O[\lambda_J^4]
\label{zeta_sp_xx}
\end{eqnarray}
In each of the  above two  Eqs. (\ref{zeta_os_xx}) and (\ref{zeta_sp_xx}), the first term on
 the right-hand side represents the COP of the uncoupled system. To compare the effects
of coupling in COPs, we calculate the difference between  $ \zeta^{\rm sp}$ and $\zeta^{\rm os}$, which is given by
\begin{eqnarray}
 \zeta^{\rm sp}-\zeta^{\rm os}=\frac{\left(T_c{\rm csch}\left[\frac{\omega}{T_h}\right]+T_h{\rm csch}\left[\frac{\omega'}{T_c}\right]\right)\lambda_J^2}{\gamma'}>0\nonumber\\
 \label{zeta_os-sp}
\end{eqnarray}
This means that a small coupling ($\lambda_J$) introduced between the subsystems makes the coupled
 spins more efficient than the coupled oscillators in transferring heat 
from cold bath to hot bath, whereas
a reverse effect is observed in the case of the heat engine ($\eta^{\rm sp}<\eta^{\rm os}$) 
for small values of $\lambda_J$.
An example is given in Fig. \ref{fig_cop}. When both the subsystems work as refrigerators, 
the global COP is bounded by the COPs of the independent systems. 
When $\lambda_J=(\omega T_c-\omega' T_h)/(T_h-T_c)=\lambda_c'$ (say), the refrigerator $A$ 
attains Carnot value with no heat transfer from the cold reservoir to the hot reservoir. Therefore, the global COP reaches its lower bound.
Further, for $\lambda_J>\lambda_c'$, the system $A$ does not work as a refrigerator 
and hence global COP is bounded from above by $\zeta_B$.
\subsection{XY model}\label{sec5B}
\begin{figure}[h]
\begin{center}
\vspace{0.2cm}
\includegraphics[width=8cm]{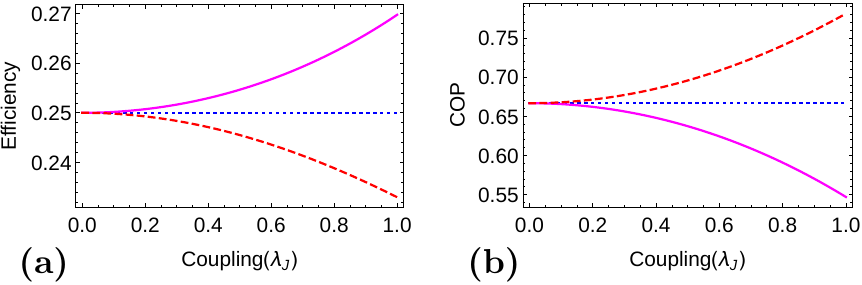}
\caption{(a) The continuous and the dashed curves represent the efficiencies 
of coupled oscillators and coupled spins respectively. The efficiency of 
the uncoupled oscillator (or spin) given in Eqs. (\ref{single_osci_eff}) and (\ref{single_spin_eff}) is shown by the 
horizontal dotted line. The parameter values are $T_h=2$, $T_c=1$, $\omega=4$ and $\omega'=3$.
(b) The COPs of coupled oscillators and spins are shown by continuous and 
dashed  curves respectively. The horizontal dotted line represents COP of the uncoupled 
oscillator (or spin) given in Eq. (\ref{single_osci_spn_zeta}). Here we used $T_h=2$, $T_c=1$, $\omega=5$ and $\omega'=2$.}
\hfill
\label{zeta_xy_fig}
\end{center}
\end{figure}

Now we consider the COPs in the case of XY model as described in Section \ref{sec4B}. 
In this model, for coupled oscillators, we have $\omega_A=\omega_B= \sqrt{\omega^2-\lambda_J^2}$
and $\omega_A'=\omega_B' =\sqrt{\omega'^2-\lambda_J^2}$.
Since both the independent oscillators are identical and having
equal frequencies at various stages of  the cycle, the COPs of the subsystems 
are equal to the global COP, which is given by
\begin{eqnarray}
 \zeta^{\rm os}&=&\frac{1}{\sqrt{\frac{\omega^2-\lambda_J^2}{\omega'^2-\lambda_J^2}}-1}.
 \label{zeta_os_xy}
\end{eqnarray}
Similarly, in the case of coupled spins, we have $\omega_A=\omega_B= \sqrt{\omega^2+\lambda_J^2}$
and $\omega_A'=\omega_B'= \sqrt{\omega'^2+\lambda_J^2}$. Both the independent 
subsystems have same energy level spacings and hence the subsystems work as refrigerators
when $\omega_A/T_h>\omega_A'/T_c$. Therefore the COPs of the subsystems are equal 
to the global COP, given by
\begin{eqnarray}
 \zeta^{\rm sp}&=&\frac{1}{\sqrt{\frac{\omega^2+\lambda_J^2}{\omega'^2+\lambda_J^2}}-1}.
 \label{zeta_sp_xy}
\end{eqnarray}
Therefore, we have  $\zeta^{\rm sp}> \zeta^{\rm os}$ even for larger values of 
$\lambda_J$. On the contrary, from Eqs. (\ref{eta_os_xy}) and (\ref{eta_sp_xy}),
we have $\eta^{\rm sp}< \eta^{\rm os}$ for the engine.
Now we can expand the COPs of the coupled systems for small values of $\lambda_J$.
\begin{equation}
 \zeta_{\rm os}=\frac{\omega'}{(\omega-\omega')}-\frac{(\omega+\omega')\lambda_J^2}{2\omega\omega'(\omega- \omega')}+O[\lambda_J^4],
\end{equation}\begin{equation}
 \zeta_{\rm sp}=\frac{\omega'}{(\omega-\omega')}+\frac{(\omega+\omega')\lambda_J^2}{2\omega\omega'(\omega- \omega')}+O[\lambda_J^4].
\end{equation}
Since $\omega>\omega'$, by introducing the coupling, the COP  of the coupled spins 
becomes higher than that of of the uncoupled spins, $\zeta_{\rm uc}=\omega'/(\omega-\omega')$ [Eq. (\ref{single_osci_spn_zeta})], 
while the COP of oscillators lies below
$\zeta_{\rm uc}$. It is interesting to note that exactly the opposite behavior is observed in the case of heat engine [see Eqs. (\ref{xy_osc_eta_exp}) and (\ref{xy_spin_eta_exp})]. 
This behaviour is  shown in Fig. \ref{zeta_xy_fig}.
\section{Discussion and future direction}\label{sec6}
To conclude, we analyzed the performances of coupled oscillators and coupled 
spins  as  heat engines and refrigerators. 
In both the cases, we choose suitable co-ordinate transformation to get two 
independent subsystems. The global figure of merit is bounded by the figures of merit
of the independent subsystems. When the upper bound is Carnot value, 
the global figure of merit reaches the lower bound. For the case of the heat engine, when one of the subsystems 
works as a refrigerator, the global efficiency falls below the lower bound.
Similarly, for the case of the refrigerator, the global COP falls below the lower bound when one of the subsystems work as a heat engine.
The upper bound is tighter than the Carnot bound.
For the systems considered in this work, it is also shown that the optimal work extractable from the coupled system 
is upper bounded by the optimal work extractable from the uncoupled system. 
Whether this is a generic feature for arbitrary coupled systems with quadratic coupling is a question of future interest. 
However, the figures of merit such as efficiency or COP can have higher values in the presence of coupling. 
We point out the range of parameters and  some forms 
of interactions where the efficiency of the coupled oscillators is higher than that of the coupled spins, whereas 
the global COP is higher for coupled spins compared to coupled oscillators.
Therefore, coupling causes opposite effects in the figures of merit of heat engine and refrigerator.

An interesting query is that whether the entanglement between the 
coupled systems is responsible for the gap in the efficiencies, i.e., the difference
between the efficiencies obtained
from coupled oscillators and coupled spins. Our preliminary observation shows that the entanglement
does not have any role in creating this gap. On the other hand, we have shown that the quantum correlations 
do not help in extracting optimal work. A recent study in optomechanical heat engine 
shows the reduction in extractable work due to quantum correlations \cite{Zhang2017}. 

One may try to look at the efficiency $\eta_1^{\rm sp}$ of coupled system of two spin-1 
particles to look
for a possible ordering of the form $\eta_{1/2}^{\rm sp}\leq\eta_1^{\rm sp}\leq\eta^{\rm os}$, 
where $\eta_{1/2}^{\rm sp}$ is efficiency in the case of a coupled spin-1/2 particles.
In the case of coupled qutrits  with $XX$ coupling, the  independent spins 
do not have
energy level spacings $\omega\pm\lambda_J$ unlike in the case of coupled spin-1/2 systems 
and coupled oscillators as discussed in Sec. \ref{sec4A} \cite{Private}. So the efficiency of the coupled 
qutrits may not exactly fall in between the efficiency curves obtained for coupled spin-1/2 
systems and coupled oscillators.  
Therefore,
an interesting  future direction is  to investigate the form of interaction for  which a
monotonic behavior  of the efficiency with the increment in the dimension of the system is exhibited.

\begin{figure}[ht]
\begin{center}
\vspace{0.2cm}
\includegraphics[width=7cm]{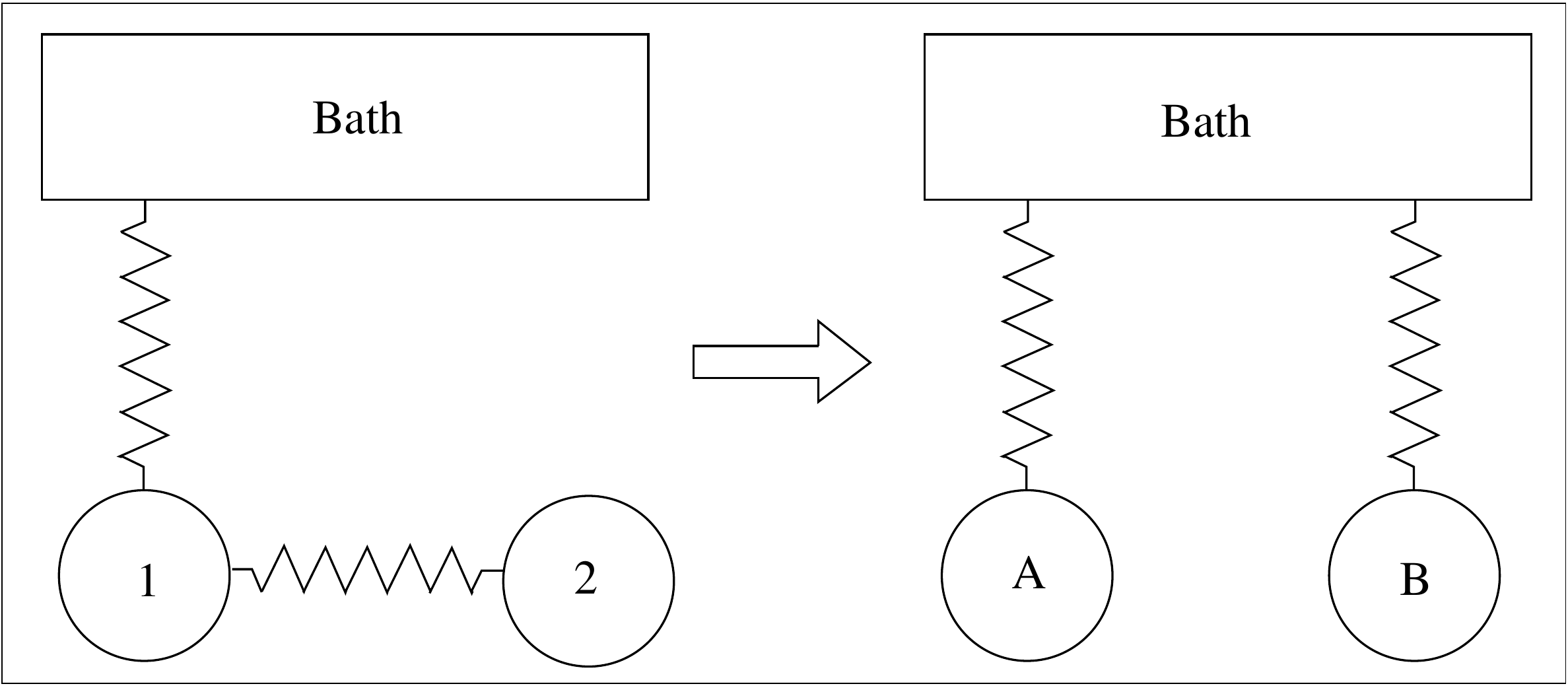}
\caption{Pictorial representation of thermalization of coupled system.}
\hfill
\label{therm}
\end{center}
\end{figure}

As mentioned in Sec. \ref{sec2}, the adiabatic processes should be done slow enough to
 avoid creation of coherence between the eigenstates of the Hamiltonian.
These coherences create a frictional effect in the system, which in turn reduces the extractable work. 
The second assumption is that the system is attached to the respective
baths for long enough time till the system attains equilibrium. 
Thermalization of coupled oscillators can be modeled with coupled micro-cavities as 
described in Ref. \cite{Zoubi'2000}. 
In this model,  one of the system (say system 1) is attached to a bath. Since the 
system 2 is coupled to the system 1,
the total system can equilibrate with the bath. When we diagonalize the Hamiltonian 
of the coupled system, the independent systems $A$ and $B$
appear to be connected with the bath (see Fig.\ref{therm}). This method is applicable even for very strongly coupled systems.

Consider an example where the first oscillator of the coupled oscillators with $XX$ interaction (see Section \ref{sec4A}), is connected 
to a bath of harmonic oscillators, characterized by an inverse temperature $\beta$. 
The  Hamiltonian for the bath and the system-bath interaction  are respectively given as \cite{Zoubi'2000}
\begin{eqnarray}
H_{B}= \sum_i(b_i^{\dagger}b_i+1/2)\omega^b_i, \;\;\;\; H_{\rm SB}= (c_1+c_1^{\dagger})\sum_i g_i(b_i+ b_i^{\dagger}),
\end{eqnarray}
where $b_i$ and $b_i^{\dagger}$ are the ladder operators for the $i$th oscillator of the bath with frequency $\omega^b_i$
and $\{g_i\}$ are the coupling strengths. So the total Hamiltonian of the system and the bath
is given as $ H^{\rm tot}=H^{\rm os}+H_{\rm B}+H_{\rm SB}$.
In the XX model, the unitary operation corresponding to the phase space transformation (given in Eq. (\ref{transformation})) is 
$c_A=(c_1+c_2)/\sqrt{2}$ and $c_B=(c_1-c_2)/\sqrt{2}$. 
Therefore, in terms of the independent modes, the total Hamiltonian can be written as
\begin{eqnarray}
 H^{\rm tot}&=&(c_A^{\dagger}c_A+1/2)\Omega_A+
 (c_B^{\dagger}c_B+1/2)\Omega_B+\sum_i(b_i^{\dagger}b_i+1/2)\omega^b_i\nonumber\\
&+&\left[(c_A+c_A^{\dagger})+(c_B+c_B^{\dagger})\right]\sum_i \frac{g_i}{\sqrt{2}}(b_i+ b_i^{\dagger})
\end{eqnarray}
Here it appears that  both the modes are independently connected to the same bath. 
This will lead to a master equation  given in  Ref. \cite{Zoubi'2000}. Further by taking the evolution of
$N_k(t)=\langle c_k^{\dagger}c_k\rangle$ of the $k$th mode (k= A,B), we get \cite{Zoubi'2000, Wang'2015}.
\begin{equation}
 N_k(t)=(N_k(0)-N_k^{\rm eq})e^{(-\Gamma_{k}t)}+N_k^{\rm eq}
\end{equation}
$\Gamma_A=\Gamma_B=\Gamma$ is the damping rate for a \textit{flat} reservoir and  $N_k^{\rm eq}=1/[\exp(\beta\Omega_k)-1]$.
This shows that the heat absorbed or rejected by the independent modes can be addressed 
separately even in finite-time case.
A similar approach is possible for the spin systems also. 

Hence an interesting direction of enquiry is to study the finite-power characteristics of the engine
which is important from the practical point of view also \cite{future}. 
This can be achieved by considering the system being in contact with the reservoir
only for a finite-time. In that case, one should also consider  fast adiabatic 
branches in the cycle. So in the adiabatic branches, the system will
create coherence due to fast changes of the external parameter and the thermalization 
processes suppresses theses coherence.
Now let us consider the time scale $\tau_{adi}$ for quantum adiabatic processes  (Stage 2 and Stage 4),
is to be much smaller than the time scales $\tau_{rel}^h$ and $\tau_{rel}^c$ that the system is attached to the hot bath  and the cold bath respectively. 
In this approximation, we can define finite power
for the engine without creating coherences between the eigenstates of the Hamiltonian during the adiabatic processes.
Therefore, for finite-time cycles, we can show the existence of the bounds for the figures of merit, as given in 
Eqs. (\ref{eta_inequalities}) and (\ref{zeta_inequalities})
because the total heat absorbed or rejected is the sum of the 
contributions from the independent subsystems.
Another possibility is to study the performance of a hybrid-system, where spin and 
harmonic oscillator are coupled. 
The Hamiltonian for such system is given as \cite{Jaynes1963,Murao1995}
\begin{equation}
 H_{hyb}=(S^{+} S^-
 +\frac{1}{2})\Omega+\left(c^{\dagger} c+\frac{1}{2}\right)\Omega+g(S^+c+S^-c^{\dagger}),
\end{equation}
where $g$ is the coupling parameter. 
This system is studied as a heat engine in a recent work \cite{Song2016}.

{\bf Acknowledgments}: GT and SG thankfully acknowledge fruitful discussions with Subhashish Banerjee, 
H. S. Mani and Ramandeep S. Johal.  The authors would like to thank the
anonymous reviewers for useful comments.



\end{document}